\documentclass[pre,twocolumn,amsmath,amssymb,floatfix,superscriptaddress]{revtex4-1}

\usepackage{graphicx}
\usepackage{color}

\begin{document}

\title{Stylized facts in social networks: Community-based static modeling} 
\author{Hang-Hyun Jo}
\affiliation{Asia Pacific Center for Theoretical Physics, Pohang 37673, Republic of Korea}
\affiliation{Department of Physics, Pohang University of Science and Technology, Pohang 37673, Republic of Korea}
\affiliation{Department of Computer Science, Aalto University School of Science, Espoo FI-00076, Finland}
\author{Yohsuke Murase}
\affiliation{RIKEN Advanced Institute for Computational Science, Kobe, Hyogo 650-0047, Japan}
\author{J\'anos T\"or\"ok}
\affiliation{Department of Theoretical Physics, Budapest University of Technology and Economics, Budapest H-1111, Hungary}
\affiliation{Center for Network Science, Central European University, Budapest H-1051, Hungary}
\author{J\'anos Kert\'esz}
\affiliation{Center for Network Science, Central European University, Budapest H-1051, Hungary}
\affiliation{Department of Theoretical Physics, Budapest University of Technology and Economics, Budapest H-1111, Hungary}
\affiliation{Department of Computer Science, Aalto University School of Science, Espoo FI-00076, Finland}
\author{Kimmo Kaski}
\affiliation{Department of Computer Science, Aalto University School of Science, Espoo FI-00076, Finland}

\date{\today}

\begin{abstract}
    The past analyses of datasets of social networks have enabled us to make empirical findings of a number of aspects of human society, which are commonly featured as stylized facts of social networks, such as broad distributions of network quantities, existence of communities, assortative mixing, and intensity-topology correlations. Since the understanding of the structure of these complex social networks is far from complete, for deeper insight into human society more comprehensive datasets and modeling of the stylized facts are needed. Although the existing dynamical and static models can generate some stylized facts, here we take an alternative approach by devising a community-based static model with heterogeneous community sizes and larger communities having smaller link density and weight. With these few assumptions we are able to generate realistic social networks that show most stylized facts for a wide range of parameters, as demonstrated numerically and analytically. Since our community-based static model is simple to implement and easily scalable, it can be used as a reference system, benchmark, or testbed for further applications.
\end{abstract}


\maketitle

\section{Introduction}\label{sect:intro}

Characterizing the social networks is of crucial importance to understand various collective dynamics taking place in them~\cite{Albert2002Statistical, Borgatti2009Network, PastorSatorras2015Epidemic}, as exemplified by disease spreading and diffusion of innovation and opinions. In recent years, the characterization of social networks in the unprecedented detail has become possible because of the availability of a number of large-scale digital datasets, e.g., face-to-face interactions~\cite{Eagle2006Reality, Zhao2011Social, Fournet2014Contact}, emails~\cite{Eckmann2004Entropy, Klimt2004Enron}, mobile phone communication~\cite{Onnela2007Structure, Blondel2015Survey}, online forums~\cite{Hric2014Community, Eom2015Tailscope}, Social Networking Services (SNSs) like Facebook~\cite{Ugander2011Anatomy} and Twitter~\cite{Kwak2010What}, and even massive multiplayer online games~\cite{Szell2010Measuring, Szell2010Multirelational}. However, these datasets capture only a part of the entire social network, implying that any conclusions derived from such datasets cannot be simply extrapolated to the whole society. Here the entire social network indicates a comprehensive picture of human social relationships with complex community structure due to today's multiple communication channels, and can be called
a multi-channel weighted social (MWS) network. This raises a series of questions: How can one translate conclusions from partial datasets to the MWS network? More importantly, what does the MWS network look like? The first question has been investigated in terms of sampling biases~\cite{Stumpf2005Subnets, Stumpf2005Sampling, Lee2006Statistical, Torok2016What}, while the second question is largely unexplored mainly due to the lack of comprehensive datasets. 

Characteristics of the MWS network are expected to be partially reflected in the empirical findings from some aspects of the network. By collecting such findings from diverse sources, we find several commonly observed features or \emph{stylized facts} of social networks~\cite{Jackson2010Social, Murase2015Modeling, Kertesz2016Multiplex}. These include broad distributions of local network quantities~\cite{Albert2002Statistical, Onnela2007Analysis}, community structure~\cite{Fortunato2010Community}, homophily~\cite{McPherson2001Birds, Newman2002Assortative}, and intensity-topology correlations~\cite{Onnela2007Structure}, etc. More recently, geographical and/or demographic information of social networks have also been explored~\cite{Onnela2011Geographic, Palchykov2012Sex, Jo2014Spatial}, which are beyond the scope of this paper. One can find the previous efforts of modeling social networks: The global picture for social networks has been described by the Granovetter's hypothesis of ``strength of weak ties''~\cite{Granovetter1973Strength}, indicating that communities of strongly connected nodes are weakly connected to each other. This picture has been empirically confirmed~\cite{Onnela2007Structure, Pappalardo2012How} and subsequently produced with computational modeling by considering cyclic and focal closure mechanisms in tie formation~\cite{Kumpula2007Emergence, Jo2011Emergence, Murase2015Modeling}. However, it was recently suggested that communities could be overlapping~\cite{Palla2005Uncovering, Ahn2010Link} in contrast to the picture of separate communities. This overlapping behavior is mostly due to the multilayer nature of social networks~\cite{Kivela2014Multilayer, Boccaletti2014Structure}, in which each layer may correspond to a certain type of human relationship or context. This means that an individual can belong to one community in one layer but simultaneously to another community in another layer. Accordingly, dynamical models for multilayer, overlapping community structure have been introduced, while reproducing other stylized facts for local network quantities~\cite{Murase2014Multilayer}. There are also several other dynamical models that partially reproduce stylized facts~\cite{Toivonen2009Comparative, Papadopoulos2012Popularity, Bagrow2013Natural}.

As many models mentioned above are dynamic and evolutionary in nature, they tend to take considerable amount of computational time. For relatively simpler and easier implementation, we take an alternative approach of static modeling to reproduce the stylized facts in social networks. For our model, we randomly assign a number of communities to a given set of isolated nodes using a few reasonable assumptions such that the community size is heterogeneous, and larger communities are assigned with smaller link density and smaller link weight. As we assign communities by hand rather than grow the network by means of some link formation mechanisms, our model can be called static. We also remark that the community size distribution is an input rather than an output of our model, although it is one of stylized facts. With the above mentioned few assumptions about communities, apparently realistic social network structures are generated showing most stylized facts for a wide range of the parameter space. Furthermore, thanks to the random nature of assigning communities to nodes, we can to some extent analytically calculate various local network quantities, e.g., for the assortative mixing, local clustering coefficient, and neighborhood overlap. 

This static modeling approach of ours is comparable to other static modeling studies, which can be classified, but not exclusively, into four categories: (i) Erd\H{o}s-R\'enyi (ER) random graphs, (ii) configuration models (CMs), (iii) stochastic blockmodels (SBMs), and (iv) exponential random graph models (ERGMs). The ER random graphs~\cite{Erdos1960Evolution} are the simplest kind of static models, and its variants have been studied, such as graphons~\cite{Goldenberg2010Survey, Airoldi2013Stochastic}, weighted random graphs~\cite{Garlaschelli2009Weighted}, or ER random graphs with community structure~\cite{Seshadhri2012Community}. In the simplest form of CMs a binary network is constructed only by using the pre-determined degree sequence of nodes, without any other correlations~\cite{Newman2010Networks}. It has been extended for containing the arbitrary distributions of subgraphs~\cite{Karrer2010Random}, to weighted networks~\cite{Britton2011Weighted, Palowitch2016Continuous}, or to networks with overlapping community structure~\cite{Jin2013Extending} or with hierarchical community structure~\cite{Stegehuis2016Powerlaw}. Next, the SBM was originally suggested for the community structure, characterized by a matrix consisting of the linking probabilities within communities and between communities~\cite{Holland1983Stochastic, Faust1992Blockmodels, Snijders1997Estimation}. As the traditional SBMs are not comparable with the empirical degree heterogeneity, the degree-corrected SBMs, by which the degree heterogeneity can be properly considered, have been studied~\cite{Karrer2011Stochastic, Fronczak2013Exponential}. The SBMs have also been extended to incorporate the overlapping communities by considering the mixed membership~\cite{Airoldi2008Mixed, Gopalan2013Efficient} or to the weighted networks, for which see Ref.~\cite{Aicher2013Adapting} and references therein. Finally, the family of ERGMs has been extensively studied in social sciences~\cite{Robins2007Introduction, Goldenberg2010Survey} as well as in terms of statistical mechanics~\cite{Park2004Statistical}. Here an ensemble of networks with given network features is considered according to the probability in the form of Boltzmann factor. The ERGMs have been extended for weighted networks~\cite{Wilson2017Stochastic} or for networks with community structure~\cite{Fronczak2013Exponential}. 

In our work, we will be exploring a different static modeling approach by explicitly considering the communities with various sizes, linking probabilities, and link weights. This way we arrive at a simple and scalable static model, which may serve as a reference system, benchmark, or testbed for further applications. 

Our paper is organized as follows: In Section~\ref{sect:stylized}, we summarize the observed stylized facts for social networks from diverse sources. Then we introduce the community-based static model in Section~\ref{sect:model}. In Section~\ref{sect:numerical}, by performing large-scale numerical simulations, we find a wide range of the parameter space in which the stylized facts are reproduced. In Section~\ref{sect:analysis}, we present the analytical results for local network quantities. Finally, we conclude our work in Section~\ref{sect:conclusion}.

\section{Stylized facts}\label{sect:stylized}

In Table~\ref{table:summary} we present a summary of the commonly observed features or \emph{stylized facts} in many digital datasets for social networks. These include the statistics of local network quantities and results for the global structure, both of which can be either topological or intensity-related. 

\begin{table*}[!ht]
    \caption{Stylized facts derived from various datasets with the expected behaviors for the MWS network~\cite{Murase2015Modeling, Kertesz2016Multiplex}. The symbol $\nearrow$ ($\searrow$) implies that the overall trend of a quantity is monotonically increasing (decreasing) according to its argument. The initially increasing and then decreasing behavior is denoted by $\nearrow \searrow$. For the Granovetterian community structure, see the main text for the details.}
  \label{table:summary}
  \begin{tabular}{ @{\hspace{0.5em}} l @{\hspace{0.5em}} @{\hspace{0.5em}} l @{\hspace{0.5em}} @{\hspace{0.5em}} l @{\hspace{0.3em}}}
      \hline 
      Category & Property or measure & Stylized fact (expectation) \\ \hline
    Topological & Degree distribution, $P(k)$ & $\searrow$ ($\nearrow \searrow$) \\
    & Average degree of neighbors as a function of degree, $k_{\rm nn}(k)$ & $\nearrow$ \\
    & Local clustering coefficient as a function of degree, $c(k)$ & $\searrow$ \\
    & Community size distribution, $P(g)$ & $\searrow$ \\
    \hline 
    Intensity-related & Strength distribution, $P(s)$ & $\searrow$ ($\nearrow \searrow$) \\
    & Weight distribution, $P(w)$ & $\searrow$ \\
    & Strength as a function of degree, $s(k)$ & $\nearrow$ \\
    & Neighborhood overlap as a function of weight, $o(w)$ & $\nearrow$ \\
    & Granovetterian community structure, $\Delta f_c$ & $> 0$ \\
    \hline
  \end{tabular}
\end{table*}

Let us first consider topological quantities. The degree $k$ of a node is the number of its neighbors. Degree distributions $P(k)$ in most datasets are found to be broad and overall decreasing~\cite{Albert2002Statistical, Onnela2007Analysis, Ugander2011Anatomy, Szule2014Lost, Eom2015Tailscope}. This implies that the most probable degree or the mode of $P(k)$, denoted by $m_k$, is of the order of $1$, leading to the fact that $m_k$ is much smaller than the average degree $\langle k\rangle$. This stylized fact is however clearly not consistent with our common sense that it is unlikely to find a majority of individuals with only one or few relationships in a society. This discrepancy can be explained by the channel selection mechanism~\cite{Torok2016What}: Most datasets for social networks represent only one communication channel or even a part of it, while social interactions generally take place on multiple communication channels, namely, the multiplexity of social networks~\cite{Kertesz2016Multiplex, Boccaletti2014Structure, Kivela2014Multilayer}. Since individuals with low degree in one channel can have high degree in another channel, by considering the multi-channel feature of social networks, $P(k)$ is expected to have a peak at the degree larger than $1$, implying that $m_k$ is of the comparable order of $\langle k\rangle$~\cite{Torok2016What, MacCarron2016Calling}. Such $P(k)$ can be called overall peaked, see Fig.~1(a) in Ref.~\cite{Galesic2012Social} for example. This does not necessarily exclude the broadness of $P(k)$.

Homophily, the tendency that alike people are attracted to each other, is one of the governing principles of social network formation~\cite{McPherson2001Birds}. One manifestation of homophily is degree assortativity, i.e., the tendency that high degree nodes are linked together. This correlation has been quantified in terms of the assortativity coefficient $\rho_{kk}$, which is a Pearson correlation coefficient (PCC) between degrees of neighboring nodes~\cite{Newman2002Assortative}. Many social networks are found to be assortative with $\rho_{kk}\approx 0.09$--$0.4$~\cite{Newman2003Structure}. This is also shown by the increasing behavior of the average degree of neighbors for nodes with degree $k$~\cite{Onnela2007Analysis, Ugander2011Anatomy}. This quantity is denoted by $k_{\rm nn}(k)$ and is defined as follows:
\begin{eqnarray}
    k_{\rm nn}(k) &\equiv & \langle k_{i,\rm nn}\rangle_{\{i|k_i=k\}},\\
    k_{i,\rm nn} &\equiv & \tfrac{1}{k_i}\sum_{j\in\Lambda_i} k_j,
\end{eqnarray}
where $\Lambda_i$ denotes the set of $i$'s neighbors and $k_i=|\Lambda_i|$ is the node $i$'s degree.

High clustering is evident in social networks, implying that your neighbor's neighbor is likely to be also your neighbor. For a node $i$, the local clustering coefficient $c_i$ is defined as the number of links between $i$'s neighbors, denoted by $e_i$, divided by the possible maximal number of links between them, i.e., $\frac{k_i(k_i-1)}{2}$, as follows:
\begin{equation}
    c_i \equiv \tfrac{2e_i}{k_i(k_i-1)}.
    \label{eq:ci_define}
\end{equation}
Then one can measure the average local clustering coefficient for nodes with degree $k$ as
\begin{equation}
    c(k) \equiv \langle c_i\rangle_{\{i|k_i=k\}}.
\end{equation}
The quantity $c(k)$ is found to be a decreasing function of $k$~\cite{Onnela2007Analysis, Szell2010Measuring, Ugander2011Anatomy, Noka2016Comparative}. This finding can be explained by considering the case that the effect of making new neighbors, corresponding to $\sim k_i^2$ in Eq.~(\ref{eq:ci_define}), is typically stronger than that of finding new links between neighbors, in relation to $e_i$ in Eq.~(\ref{eq:ci_define}). For example, if every new neighbor of a node $i$ creates a new link to one of node $i$'s existing neighbors, then $e_i\sim k_i$, leading to $c(k)\sim k^{-1}$. This behavior can be measured in terms of the PCC between $c_i$ and $k_i$, which is denoted hereafter as $\rho_{ck}$. 

At the larger scale, social networks have rich community structure: Nodes in communities are densely connected, while nodes between different communities are sparsely connected~\cite{Fortunato2010Community}. It has been shown that the community size distribution $P(g)$ has a heavy tail or power-law form, e.g., as evidenced in Refs.~\cite{Palla2005Uncovering, Ahn2010Link, Liu2012Social, Grabowicz2012Social, Hric2014Community}. We remark that in our model $P(g)$ will be used as an input rather than as an output of our model.

Next, we consider the intensity-related quantities, i.e., strength $s$ and weight $w$. The weight of a link quantifies the interaction activity between two nodes~\cite{Barrat2004Architecture}, e.g., the frequency of contact in communication. The strength of a node, denoting the activity of the node, has been defined as the sum of weights of links involving the node: 
\begin{equation}
    s_i\equiv \sum_{j\in \Lambda_i} w_{ij},
\end{equation}
where $w_{ij}$ is the weight of the link $ij$. The distributions of these quantities, $P(s)$ and $P(w)$, are also found to be broad and overall decreasing, implying that both individual and interaction activities are heterogeneous~\cite{Onnela2007Analysis}. The overall decreasing behavior of $P(w)$ can be interpreted as the prevalence of weak links~\cite{Csermely2006Weak} in social networks. 

In addition, the average strength of nodes as a function of the degree, $s(k)$, is found to be overall increasing~\cite{Onnela2007Analysis, Barrat2004Architecture}. Here $s(k)$ is defined as
\begin{equation}
    s(k) \equiv \langle s_i\rangle_{\{i|k_i=k\}}.
\end{equation}
This behavior can be measured in terms of the PCC between $s_i$ and $k_i$, denoted by $\rho_{sk}$. Combining the peaked $P(k)$ and the increasing $s(k)$, $P(s)$ is expected to be also peaked. The overall peaked $P(s)$ can be tested in terms of the mode of $P(s)$, denoted by $m_s$.

These intensities are correlated with topological properties, which can be called intensity-topology correlation or weight-topology correlation as used in Ref.~\cite{Onnela2007Structure}. A link-level consequence of weight-topology correlation can be measured by the average neighborhood overlap for links with weight $w$, denoted by $o(w)$. The neighborhood overlap of a link is the fraction of common neighbors of neighboring nodes, say $i$ and $j$, among all neighbors of those nodes:
\begin{eqnarray}
    o(w) &\equiv & \langle o_{ij}\rangle_{\{ij|w_{ij}=w\}},\\
    o_{ij} &\equiv & \tfrac{e_{ij}}{k_i+k_j-2-e_{ij}},
    \label{eq:o_ij}
\end{eqnarray}
where $e_{ij}$ denotes the number of common neighbors of nodes $i$ and $j$, whose degrees are $k_i$ and $k_j$, respectively. It has been empirically found that the stronger links tend to show larger neighborhood overlap~\cite{Onnela2007Structure, Szell2010Measuring, Takaguchi2011Predictability, Pajevic2012Organization, Koroleva2012Tie, Noka2016Comparative}, implying that closer friends tend to have more common friends. We also note that $o(w)$ begins to decrease for very large $w$ in some cases~\cite{Onnela2007Structure, Noka2016Comparative}, and that the overall decreasing $o(w)$ is found in some collaboration networks~\cite{Ke2014Tie, Noka2016Comparative}. The overall increasing behavior of $o(w)$ can be measured in terms of the PCC between $o_{ij}$ and $w_{ij}$, and it is denoted by $\rho_{ow}$.

The weight-topology correlation emerges at the global scale such that communities of strongly connected nodes are weakly connected to each other~\cite{Onnela2007Structure, Pappalardo2012How}. This is a consequence of the Granovetter's hypothesis of ``strength of weak ties''~\cite{Granovetter1973Strength}. As the Granovetterian structure is maintained by weak links, the network will be disintegrated by removing weak links rather than by removing strong links. Precisely, a link percolation analysis can be applied such that links are removed one by one either from the weakest links (ascending link removal) or from the strongest links (descending link removal) to see when the network gets disintegrated. In both cases, we may observe the percolation transition at some value of the fraction of removed links $f$. If the percolation threshold for ascending (descending) link removal is denoted by $f_c^{\rm a}$ ($f_c^{\rm d}$), we expect that 
\begin{equation}
    \Delta f_c\equiv f_c^{\rm d}- f_c^{\rm a}
\end{equation}
is significantly larger than $0$ as shown in Ref.~\cite{Murase2014Multilayer}. It is because the Granovetterian structure will be disintegrated earlier for ascending link removal than for descending link removal. Note that $\Delta f_c>0$ does not guarantee the Granovetterian structure~\cite{Garlaschelli2009Weighted}. The percolation threshold can be determined by measuring the fraction of the largest connected component (LCC) $R_{\rm LCC}$ and the susceptibility $\chi$ as functions of $f$. Here the susceptibility for the network size $N$ is defined as $\chi=\frac{1}{N}\sum_s n_s s^2$, where $n_s$ is the number of connected components of size $s$ and the summation is taken over all connected components but the LCC. At the percolation threshold $f_c$, $R_{\rm LCC}$ vanishes and $\chi$ diverges in the thermodynamic limit~\cite{Stauffer1994Introduction}. 

Finally, we remark that these stylized facts have been deduced mostly from datasets of single communication channels, while here we are interested in the multi-channel weighted social (MWS) network representing a multiplicity of communication channels. Empirical findings in single channel datasets may reflect some properties of the reality, but they may also introduce biases~\cite{Torok2016What}. Thus, our aim in this paper is to reproduce, in addition to the stylized facts, the expected peaked behavior of $P(k)$ and $P(s)$ for the MWS network, as listed in Table~\ref{table:summary} with expected behavior in parenthesis.

\section{Model}\label{sect:model}

\begin{figure}[!t]
    \includegraphics[width=.9\columnwidth]{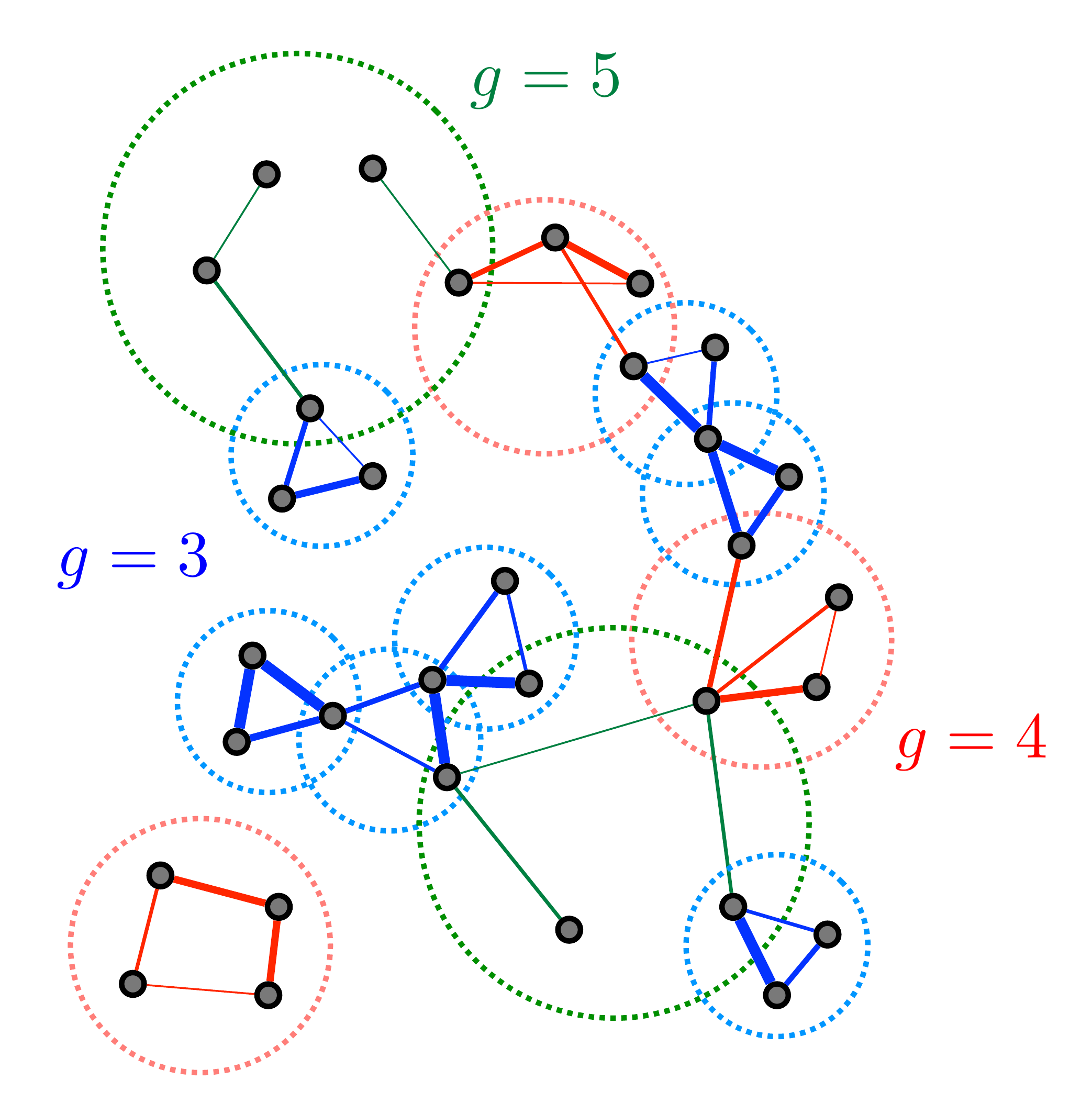}
    \caption{(Color online) Schematic diagram of the community-based static model with overlapping communities of various sizes with $g=3$, $4$, and $5$. The larger dotted circle indicates the larger community size, and thicker lines indicate larger link weights.}
    \label{fig:visualNet}
\end{figure}

In order to devise a community-based static model for stylized facts in social networks, let us consider an undirected weighted network of $N$ nodes and $C$ communities. For each community, the community size $g$ is drawn from a distribution $P(g)$. The minimum size of the community is set as $g_0=3$, i.e., a triangle. The maximum size of the community $g_{\rm max}$ is set to be $10^3$ that is sufficiently large but not too large for being realistic. Then $g$ different nodes are randomly chosen to form a community, where links between them are created with linking probability $p(g)$. Here $p(g)$ is a decreasing function of $g$ to represent the tendency of link density being sparser for larger communities. Each of these links is assumed to have a positive weight randomly drawn from a distribution $P_g(w)$ with a characteristic weight $w(g)$. Here $w(g)$ is a decreasing function of $g$ to represent the tendency of weaker ties in larger communities. For these reasons, we assume the following functional forms in the range of $g_0\leq g\leq g_{\rm max}$:
\begin{eqnarray}
    P(g)&\equiv & Ag^{-\alpha},\\
    \label{eq:model_pg}
    p(g)&\equiv & \left(\tfrac{g_0}{g}\right)^\beta,\\
    \label{eq:model_wg}
    P_g(w)&\equiv &\tfrac{\theta[w(g)-w]}{w(g)},\ w(g)\equiv w_0 \left(\tfrac{g_0}{g}\right)^\gamma
\end{eqnarray}
with non-negative exponents $\alpha$, $\beta$, and $\gamma$. Here $A\equiv (\sum_{g=g_0}^{g_{\rm max}} g^{-\alpha})^{-1}$ denotes a normalization constant, and $\theta(\cdot)$ is the Heaviside step function. We take a power-law form for $P(g)$ based on the empirical findings in Refs.~\cite{Palla2005Uncovering, Ahn2010Link, Liu2012Social, Hric2014Community}. The power-law form for $p(g)$ is also inspired by the observations~\cite{Hric2014Community}. Finally, we take the uniform distribution for $P_g(w)$ in the range of $(0,w(g)$$]$ as no evidence is known for its shape. Other candidates for $P_g(w)$ will be shortly discussed in Appendix~\ref{appendix:Pgw}. These settings immediately imply that the average number of neighbors due to the community of size $g$ is $(g-1)p(g)$, i.e., of the order of $g^{1-\beta}$. Since a membership to the larger community could imply more ties created, it is assumed that $\beta \leq 1$. Further, the sum of weights due to those neighbors in the community of size $g$ scales with $g^{1-\beta-\gamma}$. This sum decreases with $g$ for sufficiently large $\gamma$, meaning that people may spend less time in communication with the members of the larger communities. See Fig.~\ref{fig:visualNet} for the schematic diagram.

Since in social networks each node may belong to multiple communities, the network has an overlapping community structure and this is assured here by construction. Moreover, such network can be interpreted in the frame of a multiplex network~\cite{Kivela2014Multilayer, Boccaletti2014Structure}. In our model each \emph{layer} indexed by $g$ is defined as a set of links created in communities of the same size $g$~\footnote{Note that the layers in the real social networks do not have to consist of the communities of the same size.}. Then, a pair of nodes may be connected by multiple links when they belong to multiple communities, irrespective of whether those communities are in the same layer or not. However, we reduce multiplex weighted networks generated by our model to single layer weighted networks by assigning a unique weight to each pair of nodes. Here we use the rule of weight aggregation such that a weight $w_{ij}$ for a link $ij$ is given by
\begin{equation}
    w_{ij}\equiv \sum_{g\in G_{ij}} w_{ij,g},
\end{equation}
where $G_{ij}$ denotes a set of $g$s for communities in which the link $ij$ is created, and $w_{ij,g}$ is the weight of link $ij$ created in the layer $g$. Note that alternative rules for weight aggregation can be used, e.g., by taking a maximum weight from a set of $w_{ij,g}$.

We remark that it is very unlikely for a pair of nodes to belong to multiple communities, which is indeed asymptotically the case as shown in Appendix~\ref{appendix:overlap}. Hence, links can be divided into exclusive sets according to the community size $g$ in which the link is created, implying that $|G_{ij}|=1$ for all links. 

\section{Numerical Results}\label{sect:numerical}

For numerical simulations, we generate networks of size $N=3\cdot 10^4$ with $g_0=3$ and $g_{\rm max}=10^3$. We also set $w_0=1$ without loss of generality. We begin with $N$ isolated nodes, to which communities are added sequentially until the average degree reaches a predetermined value, e.g., $\langle k\rangle=100$. This in turn determines the total number of communities, $C$. We first study the effects of $\alpha$ and $\beta$ on the topological properties of the generated networks to find the parameter region for reproducing the stylized facts. Then, for a fixed value of $\beta$ within the region for the topological stylized facts, we study the effects of $\alpha$ and $\gamma$ on the intensity-related properties of generated networks.

\subsection{Topological properties}

\begin{figure}[!t]
    \includegraphics[width=\columnwidth]{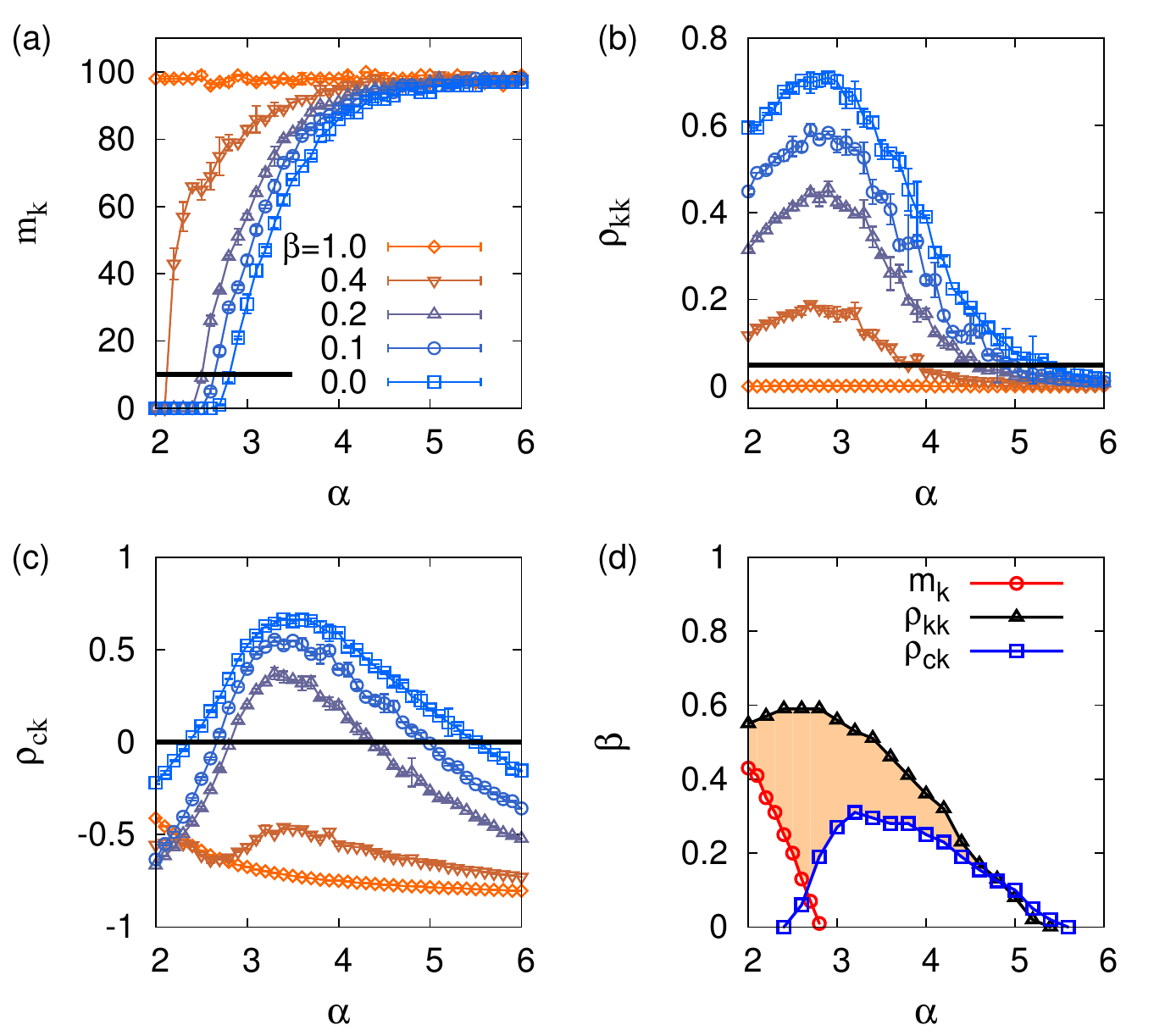}
    \caption{(Color online) Effects of $\alpha$ and $\beta$ on topological properties: (a) $m_k$, (b) $\rho_{kk}$, and (c) $\rho_{ck}$, with corresponding threshold values (black lines) for stylized facts. (d) Assuming that the topological stylized facts are reproduced when $m_k>10$, $\rho_{kk}>0.05$, and $\rho_{ck}<0$, we derive the parameter region (shaded area) surrounded by three curves from (a--c). Here the results have been averaged over $10$--$40$ networks generated using $N=3\cdot 10^4$, $\langle k\rangle=100$, $g_0=3$, and $g_{\rm max}=10^3$.}
    \label{fig:topology_alphabeta}
\end{figure}

We obtain the simulation results of topological quantities, i.e., $m_k$, $\rho_{kk}$, and $\rho_{ck}$, for the wide range of $\alpha$ and $\beta$. In Fig.~\ref{fig:topology_alphabeta}(a), $m_k$ is found to overall increase with $\alpha$ and $\beta$. As $\langle k\rangle=100$ in the simulations, we take a threshold value for $m_k$ as $10$ to distinguish the region of overall decreasing $P(k)$ for small $\alpha$ and $\beta$ from that of overall peaked $P(k)$ for large $\alpha$ and $\beta$. This threshold value is arbitrary yet reasonable for the qualitative description. In Fig.~\ref{fig:topology_alphabeta}(b), we find that $\rho_{kk}\geq 0$ for the entire range of the parameter space, and that $\rho_{kk}$ overall decreases with $\alpha$ and $\beta$. Here we take $0.05$ as the threshold value to separate the assortative region for small $\alpha$ and $\beta$ from the uncorrelated region for large $\alpha$ and $\beta$. Finally, we observe the non-monotonic behavior of $\rho_{ck}$ according to $\alpha$ and $\beta$, as shown in Fig.~\ref{fig:topology_alphabeta}(c). For small $\beta$, there exists an intermediate region of $\alpha$ with $\rho_{ck}>0$. As $\beta$ increases, this region becomes narrower and finally disappears. Summarizing the results, we find the wide region of $\alpha$ spanning over $(2,4.5)$ and of $\beta$ spanning over $(0.1,0.6)$, where all topological stylized facts are reproduced, i.e., relatively large $m_k$ and $\rho_{kk}$ as well as negative $\rho_{ck}$. This region is depicted as shaded in Fig.~\ref{fig:topology_alphabeta}(d). 

In order to understand these results, we remind that the larger value of $\alpha$ leads to more frequent triangles as $g_0=3$, and that the larger value of $\beta$ leads to sparser communities, especially for relatively large communities. We first consider the limiting case with large $\alpha$ and $\beta$, and then the other limiting case with small $\alpha$ and $\beta$.

In the limiting case when $\alpha\to\infty$, all communities are triangles, irrespective of the value of $\beta$. The network is then a random graph but consisting of triangles, namely a random $3$-uniform hypergraph in graph theory~\cite{Karonski2002Phase}. As $P(k)$ is peaked~\cite{Lopez2013Distribution}, $m_k$ is much larger than $1$ for sufficiently large $\langle k\rangle$. We numerically find that $m_k\approx \langle k\rangle$ for $\alpha\geq 5$ is independent of $\beta$, as seen in Fig.~\ref{fig:topology_alphabeta}(a). It is also expected that there is no degree-degree correlations, i.e., $\rho_{kk}\approx 0$. Furthermore, the local clustering coefficient may be negatively correlated with $k$, i.e., $\rho_{ck}<0$, as explained in Section~\ref{sect:stylized}. We find similar behaviors in the limiting case when $\beta\to\infty$, as communities of size larger than $g_0$ practically have no links in them, irrespective of $\alpha$. These arguments are consistent with simulation results already when $\beta=1$, as depicted in Fig.~\ref{fig:topology_alphabeta}(a--c).

On the other hand, when $\beta=0$, we have $p(g)=1$ for the entire range of $g$, leading to the network of overlapping cliques of various sizes. If the value of $\alpha$ decreases from infinity, more and more large cliques are created in the network. Since the total number of links is fixed, links are then concentrated more in large communities, leading to a number of isolated nodes. The value of $m_k$ then decreases from $\langle k\rangle$ as $\alpha$ decreases. Degrees of nodes in the same community tend to be similar, which along with the heterogeneous community size for finite $\alpha$ leads to the positive $\rho_{kk}$. As for the local clustering coefficient $c$, it must be $1$ for all connected nodes unless communities are overlapping. Since communities overlap in general, the value of $\rho_{ck}$ is expected to be negative as nodes belonging to multiple communities tend to have higher $k$ but smaller $c$. However, in the intermediate range of $\alpha$ the positive $\rho_{ck}$ is found in Fig.~\ref{fig:topology_alphabeta}(c). It is because nodes belonging to even larger communities tend to have higher $k$ and to find more links between their neighbors, hence the larger value of $c$. If $\alpha$ becomes too small, most nodes have $c$ of the order of $1$, while some nodes with high $k$ have smaller values of $c$ due to the overlapping communities. Therefore the negative $\rho_{ck}$ is observed, while the global clustering, i.e., the average of $c$ for all nodes in the network, is still high (not shown).

\begin{figure}[!t]
    \includegraphics[width=\columnwidth]{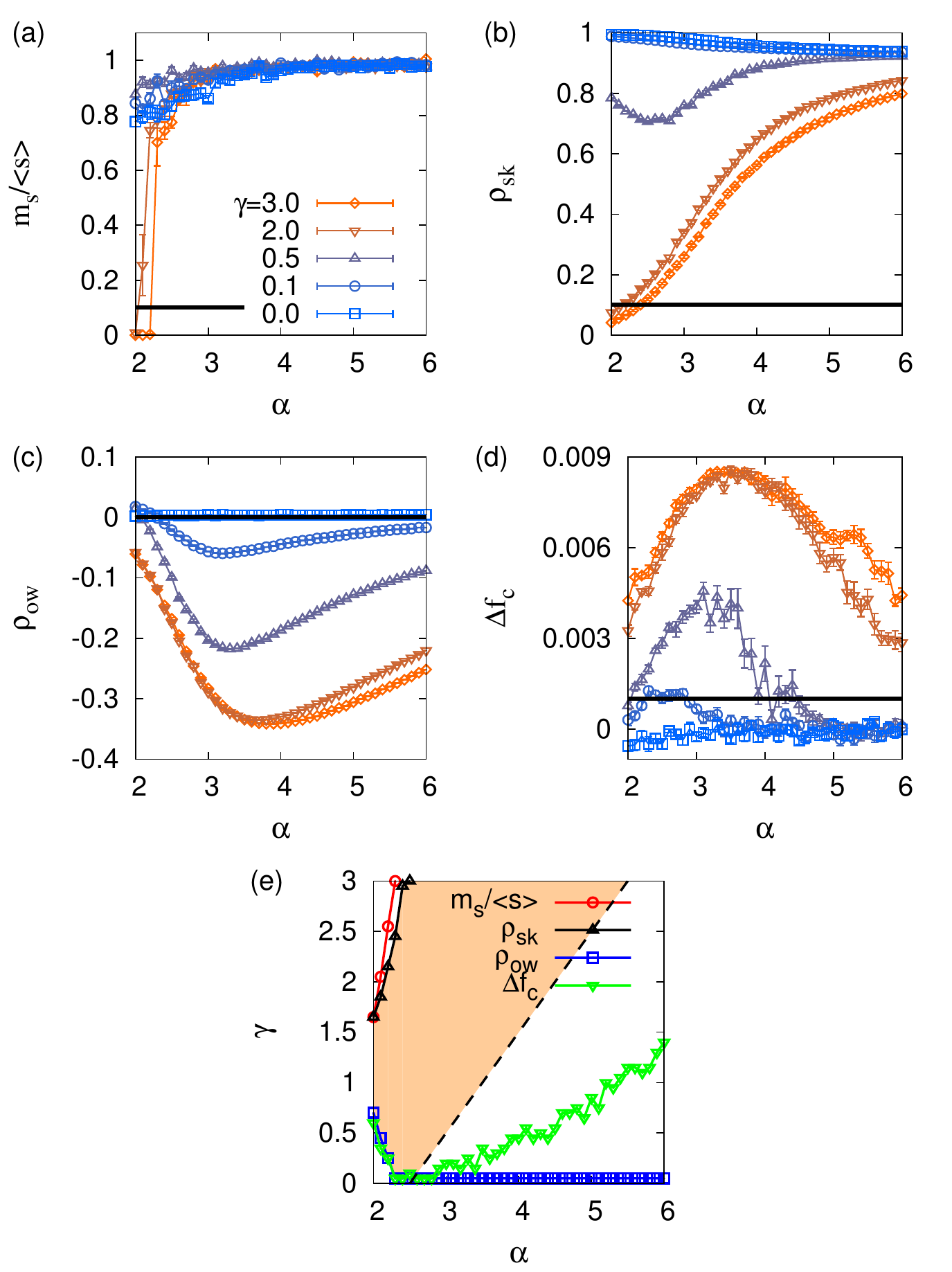}
    \caption{(Color online) Effects of $\alpha$ and $\gamma$ on intensity-related properties for a given $\beta=0.5$: (a) $\frac{m_s}{\langle s\rangle}$, (b) $\rho_{sk}$, (c) $\rho_{ow}$, and (d) $\Delta f_c$, with corresponding threshold values (black lines) for stylized facts. (e) Assuming that the intensity-related stylized facts are reproduced when $\frac{m_s}{\langle s\rangle}>0.1$, $\rho_{sk}>0.1$, $\rho_{ow}>0$, and $\Delta f_c>0.001$, we barely find the parameter region surrounded by four curves from (a--d). If the condition for positive $\rho_{ow}$ is relaxed, then we have the wide range of parameter space (shaded area) for the stylized facts, i.e., for large $\alpha$ and $\gamma$. Here the results have been averaged over $10$ networks generated using $N=3\cdot 10^4$, $\langle k\rangle=100$, $g_0=3$, $g_{\rm max}=10^3$, and $w_0=1$. The dashed line, $\gamma=\alpha+\beta-3$, indicates the criterion for the decreasing behavior of $P(w)$ obtained from Eq.~(\ref{eq:decreasingPw}).}
    \label{fig:activity_alphagamma}
\end{figure}

\subsection{Intensity-related properties}

Next we numerically obtain the intensity-related quantities, i.e., $\frac{m_s}{\langle s\rangle}$, $\rho_{sk}$, $\rho_{ow}$, and $\Delta f_c$, by varying $\alpha$ and $\gamma$ while keeping $\beta=0.5$, for which it is expected to show the topological stylized facts for $\alpha<3.5$. Here the mode of $P(s)$ has been normalized by the average strength $\langle s\rangle$ as $\langle s\rangle$ also depends on the parameters. We find in Fig.~\ref{fig:activity_alphagamma}(a) that $\frac{m_s}{\langle s\rangle}$ overall increases with $\alpha$ but decreases with $\gamma$. In order to distinguish the region of overall decreasing $P(s)$ for small $\alpha$ and large $\gamma$ from that of overall peaked $P(s)$ for large $\alpha$ and small $\gamma$, we take a threshold value for $\frac{m_s}{\langle s\rangle}$ as $0.1$ to imitate the threshold value of $m_k$. In Fig.~\ref{fig:activity_alphagamma}(b), we find that $\rho_{sk}\geq 0$ for the entire range of the parameter space, and that $\rho_{sk}$ overall increases with $\alpha$ but decreases with $\gamma$. We take $0.1$ as the threshold value for $\rho_{sk}$ to separate the correlated region for large $\alpha$ and small $\gamma$ from the uncorrelated region for small $\alpha$ and large $\gamma$. As for $\rho_{ow}$, it has a negative value for the almost entire range of the parameter space, except when both $\alpha$ and $\gamma$ are small, as shown in Fig.~\ref{fig:activity_alphagamma}(c). Finally, $\Delta f_c$ shows the non-monotonic behavior according to $\alpha$ in Fig.~\ref{fig:activity_alphagamma}(d). For small $\gamma$, there exists an intermediate region of $\alpha$ showing $\Delta f_c>0.001$, with the threshold value of $0.001$. As $\gamma$ increases, this region becomes wider. 

In summary, we barely find a region in the parameter space of $(\alpha, \gamma)$ for a fixed $\beta=0.5$, where all intensity-related stylized facts are reproduced, i.e., relatively large $\frac{m_s}{\langle s\rangle}$, $\rho_{sk}$, and $\Delta f_c$, as well as positive $\rho_{ow}$. However, if the condition for positive $\rho_{ow}$ is relaxed, we have the wide range of the parameter space for the stylized facts to be reproduced, i.e., for large $\alpha$ and $\gamma$. This region is depicted as shaded in Fig.~\ref{fig:activity_alphagamma}(e). We note that despite the negative $\rho_{ow}$, $o(w)$ shows the increasing and then decreasing behavior for some range of parameter values, which will be discussed in the next Section. 

In order to understand these results, we remind that the larger value of $\gamma$ leads to more hierarchical weights as larger communities contain even weaker links on average. We first consider the limiting case with small $\gamma$ and large $\alpha$, and then the case with large $\gamma$ and not too small $\alpha$.

In the limiting case when $\gamma=0$, all weights are drawn from the uniform distribution with the range of $[0,1]$, independent of the community size. It implies no intensity-topology correlations, irrespective of topological structure of the network. Therefore, we have $s(k)\approx \frac{k}{2}$ for sufficiently large networks, which leads to $\frac{m_s}{\langle s\rangle}\approx \frac{m_k}{\langle k\rangle}\sim \mathcal{O}(1)$ and $\rho_{sk}\approx 1$ for a wide range of $\alpha$. Since the neighborhood overlap of a link is statistically independent of its weight, we have $\rho_{ow}\approx 0$. Finally, for the link percolation the ascending link removal is just the same as the descending link removal, thus $\Delta f_c\approx 0$ is expected. These expected behaviors are numerically confirmed as shown in Fig.~\ref{fig:activity_alphagamma}(a--d). These tendencies are also observed for very large $\alpha$, independent of $\gamma$, as the network consists of triangles only and then the weights are fully uncorrelated with topological structure.

On the other hand, the large value of $\gamma$ enhances the hierarchical structure of weights. If $\alpha$ decreases from infinity, the network is dominated by larger communities, i.e., by the larger number of links with very small weights. Then as more nodes have smaller strength, both $m_s$ and $\langle s\rangle$ decrease to $0$, while $\frac{m_s}{\langle s\rangle}$ decreases to $0$ as $P(s)$ becomes more skewed due to the nodes belonging only to large communities. Since the majority of links are weak, the higher degree does not necessarily mean the larger strength, leading to the smaller $\rho_{sk}$ for decreasing $\alpha$. Further, since the nodes in the large communities tend to have more common neighbors and weaker links between them, we find negative correlations between neighborhood overlap and weight of links, except for small $\alpha$ and $\gamma$, as shown in Fig.~\ref{fig:activity_alphagamma}(c). Finally, the Granovetterian community structure appears to exist in the networks, as evidenced by $\Delta f_c>0$ for sufficiently large $\gamma$ and for the wide range of $\alpha$ in Fig.~\ref{fig:activity_alphagamma}(d). This is because by construction weak links in larger communities connect smaller communities containing strong links. Moreover, as the larger $\gamma$ enhances the hierarchical structure of weights, we find the stronger weight-topology correlation, hence a larger $\Delta f_c$. 

We have also studied the effects of $\beta$ and $\gamma$ for a fixed value of $\alpha=2.5$ to find the range of the parameter space $(\beta,\gamma)$ for the intensity-related stylized facts. See Appendix~\ref{appendix:betagamma} and Fig.~\ref{fig:activity_betagamma} for the details.

To conclude the Section, we note that our simple model with a few assumptions about communities could reproduce almost all stylized facts, except for the increasing neighborhood overlap as a function of link weight, for the wide range of the parameter space $(\alpha,\beta,\gamma)$.

\section{Analysis of local network quantities}\label{sect:analysis}

Thanks to the random nature of assigning communities to nodes in our model, we can obtain analytical solutions of the local network quantities to some extent, which is important for the rigorous understanding of the consequences of the model. Since the network consists of many random communities whose sizes are randomly drawn from a given distribution, the network can be understood in terms of an aggregate of different layers, where each layer consists of communities with the same size that are overlapping by construction. Based on this concept of layers, we first get analytical results for topological quantities, regardless of the rule of assigning weights to links: $P(k)$, $c(k)$, and $k_{\rm nn}(k)$. We then consider intensity-related quantities: $P(w)$, $P(s)$, $s(k)$, and $o(w)$. According to the discussion in Appendix~\ref{appendix:overlap}, we assume that communities do not overlap over more than one node.

\begin{figure}[!t]
    \includegraphics[width=\columnwidth]{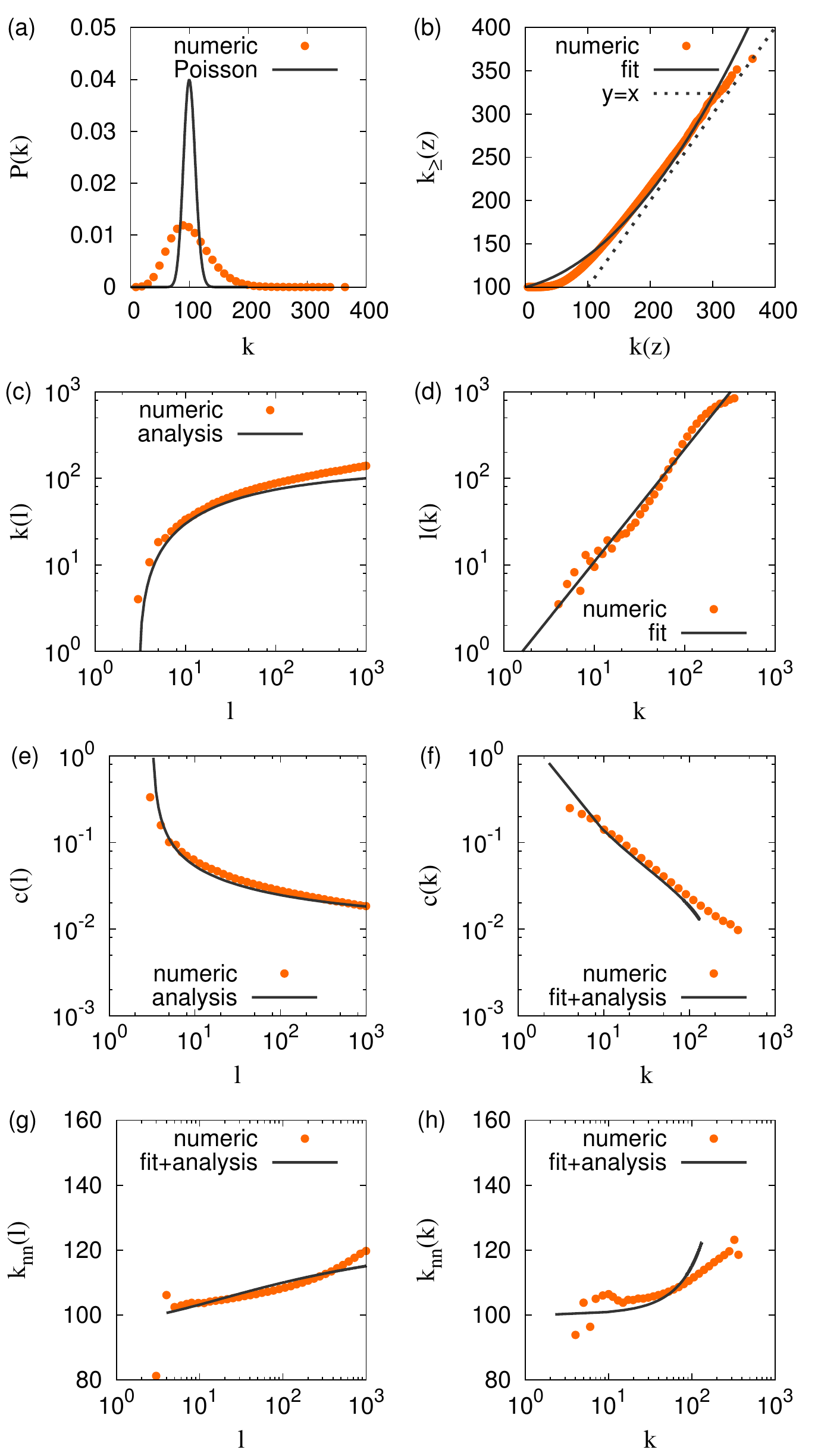}
    \caption{(Color online) Simulation results of topological quantities for $N=3\cdot 10^4$, $\langle k\rangle=100$, $g_0=3$, $g_{\rm max}=10^3$, $\alpha=2.7$, and $\beta=0.55$ (circles) are compared to the analytical results (solid lines). The simulation results were averaged over $50$ generated networks.}
    \label{fig:analysis_topology}
\end{figure}

\subsection{Topological quantities}

The number of communities of size $g$ is simply given by $n_g\equiv CP(g)=CAg^{-\alpha}$. The average number of links in the layer $g$ is equal to $L_g\equiv n_g\tfrac{g(g-1)}{2} p(g)$ and the total number of links in the network is obtained by $L=\sum_g L_g$. Then $C$ is determined from the relation $L=\frac{N\langle k\rangle}{2}$. The degree distribution in the layer $g$ can be naively approximated by
\begin{equation}
    P_g(k) = {N-1 \choose k} p_g^k (1-p_g)^{N-1-k} \approx e^{-\langle K\rangle_g} \tfrac{\langle K\rangle_g^k}{k!}
\end{equation}
with $p_g\equiv \frac{2L_g}{N(N-1)}$ and $\langle K\rangle_g \equiv \frac{2L_g}{N}$ respectively denoting the link density and the average degree in the layer $g$. As we assume that links in different layers are rarely overlapping, the degree distribution for the whole network is obtained as a Poisson distribution:
\begin{equation}
    \label{eq:Pk_simple}
    P(k)= e^{-\langle k\rangle} \tfrac{\langle k\rangle^k}{k!},
\end{equation}
where we have used the relation $\langle k\rangle = \sum_g \langle K\rangle_g$. $P(k)$ is found to have a peak around at $\langle k\rangle$, i.e., $m_k\approx \langle k\rangle$. The numerical $P(k)$ in Fig.~\ref{fig:analysis_topology}(a) also has the peak around at $\langle k\rangle$ but it is broader than the above Poisson distribution possibly due to the correlation between links imposed by the community structure. This effect due to the correlation can be explained by separating nodes chosen for communities from those not chosen in each layer. The number of nodes chosen for communities of size $g$ is approximated as $N_g\approx gn_g$, leading to $\langle k\rangle_g\equiv \frac{2L_g}{N_g}\approx (g-1)p(g)$ denoting the average degree only for nodes chosen for communities in the layer $g$. Based on this idea, we derive the analytical result for the standard deviation $\sigma$ for the degrees in Appendix~\ref{appendix:std_degree}. For example, when $\alpha=2.7$ and $\beta=0.55$, we obtain $\sigma\approx 31.6$ using Eq.~(\ref{eq:std_degree}), which is comparable with the simulation result of $\approx 34.3$. Note that the standard deviation for the Poisson distribution with $\langle k\rangle=100$ is $10$. Thus, the correlation between links imposed by the communities is important in understanding the broader $P(k)$ than the Poissonian case.

Now we focus on the egocentric network that consists of a node and its neighbors, namely, the ego and its alters. The degree of the node is determined by the layers in which the node was chosen to create links. We denote by $l$ the largest community size or the \emph{outermost layer} involving the node. Since the layers involving the node or the ego are not necessarily consecutive, we take a mean-field approach: The indexes of layers involving the ego are assumed to be consecutive from $g_0$ to $l$. Since the probability of an ego being chosen to have links in the layer $g$ is $\frac{N_g}{N}$, one obtains the expected degree of nodes with outermost layer $l$ as
\begin{eqnarray}
    \label{eq:define_k_l}
    k(l) &\equiv & \sum_{g=g_0}^{l} \tfrac{N_g}{N} \langle k\rangle_g 
    \approx \tfrac{g_0^{\beta}CA}{N} \int_{g_0}^{l} (g^{2-\alpha-\beta}- g^{1-\alpha-\beta}) dg \nonumber\\
    &=& \tfrac{g_0^{\beta}CA}{N} \left[ H^{3,1}_{1,0}(g_0,l)- H^{2,1}_{1,0}(g_0,l) \right].
    \label{eq:k_l}
\end{eqnarray}
Here we have defined for convenience
\begin{eqnarray}
    && H^{u,v}_{m,n}(x_0,x_1)\equiv \int_{x_0}^{x_1} x^{u-v\alpha-m\beta-n\gamma-1}dx\\ \label{eq:H}
    &&=\left\{\begin{tabular}{ll}
            $\tfrac{x_1^{u-v\alpha-m\beta-n\gamma} -x_0^{u-v\alpha-m\beta-n\gamma}}{u-v\alpha-m\beta-n\gamma}$ & if $u-v\alpha-m\beta-n\gamma\neq 0$,\\
            $\ln \frac{x_1}{x_0}$ & otherwise.
        \end{tabular}\right.\nonumber\\
\end{eqnarray}
Figure~\ref{fig:analysis_topology}(c) shows that this analytical result is comparable with the simulation results. If $\alpha+\beta\neq 2,\ 3$, then $k(l)$ can be explicitly written as follows:
\begin{equation}
    \label{eq:k_l_approx}
    k(l)\approx a_0 + a_1 l^{3-\alpha-\beta} + a_2 l^{2-\alpha-\beta},\\
\end{equation}
where
\begin{eqnarray}
    a_0 &\equiv& \tfrac{g_0^\beta CA}{N}\left( -\tfrac{g_0^{3-\alpha-\beta}}{3-\alpha-\beta} +\tfrac{g_0^{2-\alpha-\beta}}{2-\alpha-\beta}\right),\\
    a_1 &\equiv& \tfrac{g_0^\beta CA}{N(3-\alpha-\beta)},\
    a_2 \equiv -\tfrac{g_0^\beta CA}{N(2-\alpha-\beta)}.
\end{eqnarray}

Later we will need to use the expected outermost layer of nodes with degree $k$, denoted by $l(k)$. As the calculation of $l(k)$ is not trivial, we fit the simulation result in Fig.~\ref{fig:analysis_topology}(d) with a simple scaling function as follows:
\begin{equation}
    \label{eq:l_k_fit}
    l(k)\approx \tilde a k^{\mu}
\end{equation}
with $\tilde a\approx 0.542$ and $\mu\approx 1.30$. We remark that the inverse function of $k(l)$ does not necessarily match with $l(k)$, implying that in general $k[l(k)]\neq k$.

Next, we calculate the expected local clustering coefficient in two forms: $c(k)$ and $c(l)$. Since the probability of linking two alters in the same layer is $p(g)$, we obtain
\begin{eqnarray}
    && c(l) \equiv \tfrac{1}{k(l)[k(l)-1]} \sum_{g=g_0}^{l} \tfrac{N_g}{N} \langle k\rangle_g(\langle k\rangle_g-1)p(g)\\
    && \approx \tfrac{g_0^{3\beta}CA}{Nk(l)[k(l)-1]} \left[ H^{4,1}_{3,0}(g_0,l)-2H^{3,1}_{3,0}(g_0,l)+H^{2,1}_{3,0}(g_0,l) \right]\nonumber\\
    && -\tfrac{g_0^{2\beta}CA}{Nk(l)[k(l)-1]} \left[ H^{3,1}_{2,0}(g_0,l)-H^{2,1}_{2,0}(g_0,l) \right].
    \label{eq:lcc_k}
\end{eqnarray}
Figure~\ref{fig:analysis_topology}(e) shows that this analytical result is comparable with the simulation results. Then, we numerically get $c(k)$ by plugging $l(k)$ in Eq.~(\ref{eq:l_k_fit}) into Eq.~(\ref{eq:lcc_k}), i.e.,
\begin{equation}
    c(k)\equiv c[l=l(k)],
\end{equation}
which is plotted against $k[l(k)]$ in Fig.~\ref{fig:analysis_topology}(f). We find our approximated solution to be comparable with the simulation results.

Then, we study the assortative mixing by calculating $k_{\rm nn}(k)$ and $k_{\rm nn}(l)$. Let us consider the above egocentric network, where each alter, say $j$, has been connected to the ego in some layer, say $z$. If $z$ is large, $z$ is most likely to be the outermost layer of $j$. Then the degree of $j$ is $k(z)$ that is obtained using Eq.~(\ref{eq:k_l}). However, for small $z$, the outermost layer of $j$ is most likely to be larger than $z$. Hence the degree of $j$ can be estimated as the average degree of nodes whose outermost layer is equal to or larger than $z$. If $z$ is very small, the expected degree of $j$ would be $\langle k\rangle$. In general, the expected degree of the alter having a link to the ego in the layer $z$ can be obtained as follows:
\begin{equation}
    k_{\geq}(z) \equiv \tfrac{ \sum_{k=k(z)}^\infty kP(k)}{ \sum_{k=k(z)}^\infty P(k)}.
\end{equation}
In Fig.~\ref{fig:analysis_topology}(b), we plot $k_{\geq}(z)$, calculated using the numerical $P(k)$, against $k(z)$. As expected, $k_{\geq}(z)$ has a value of $\langle k\rangle$ for small $z$, while it approaches $k(z)$ for large $z$. As the calculation of $k_{\geq}(z)$ is not trivial, we make an approximation in a quadratic form as
\begin{equation}
    \label{eq:k_z_fit}
    k_{\geq}(z) \approx \langle k\rangle + b_1 k(z)+ b_2 k(z)^2
\end{equation}
with coefficients $b_1\approx 0.178$ and $b_2\approx 0.00185$, see the solid line in Fig.~\ref{fig:analysis_topology}(b). More complicated functional form may give a better result, but it makes further calculations much more difficult.

Since the ego has on average $\langle k\rangle_g$ neighbors in the layer $g$ whose expected degree is $k_{\geq}(g)$, using Eq.~(\ref{eq:k_z_fit}) we obtain
\begin{eqnarray}
    \label{eq:define_knn}
    && k_{\rm nn}(l) \equiv \tfrac{1}{k(l)} \sum_{g=g_0}^{l} \tfrac{N_g}{N} \langle k\rangle_g k_{\geq}(g) \\
    && \approx \tfrac{g_0^{\beta}CA}{Nk(l)} \big\{ 
        c_0[ H^{3,1}_{1,0}(g_0,l) - H^{2,1}_{1,0}(g_0,l)]
        - c_1 H^{4,2}_{2,0}(g_0,l)
        \nonumber\\ 
        && 
        + (c_1-c_2) H^{5,2}_{2,0}(g_0,l) 
        + c_2 H^{6,2}_{2,0}(g_0,l) 
        - c_3 H^{6,3}_{3,0}(g_0,l) 
        \nonumber\\
        && 
        + (c_3-c_4) H^{7,3}_{3,0}(g_0,l) 
        + (c_4-c_5) H^{8,3}_{3,0}(g_0,l) 
        + c_5 H^{9,3}_{3,0}(g_0,l)
        \big\},
        \nonumber\\
        \label{eq:knn_k_approx}
\end{eqnarray}
where
\begin{eqnarray}
    c_0 \equiv \langle k\rangle +a_0b_1+a_0^2b_2,&&\
    c_1 \equiv a_2(b_1+2a_0b_2), \nonumber\\
    c_2 \equiv a_1(b_1+2a_0b_2),&&\
    c_3 \equiv a_2^2b_2,\\
    c_4 \equiv 2a_1a_2b_2,&&\
    c_5 \equiv a_1^2b_2.\nonumber
\end{eqnarray}
Here we have assumed that $\alpha+\beta\neq 2,\ 3$. Figure~\ref{fig:analysis_topology}(g) shows that Eq.~(\ref{eq:knn_k_approx}) is comparable with the simulation results. If $\alpha+\beta<3$, as $b_1\gg b_2$,
one may expect that for very large $l$
\begin{equation}
    k_{\rm nn}(l)\sim l^{3-\alpha-\beta},
\end{equation}
which hints at the role of exponent values such that the higher assortativity is obtained with the smaller values of $\alpha$ and $\beta$, i.e., the more heterogeneous size of communities and the more links for larger communities. This is consistent with the results in Fig.~\ref{fig:topology_alphabeta}(b). Then, we numerically get $k_{\rm nn}(k)$ by assuming that
\begin{equation}
    k_{\rm nn}(k)\equiv k_{\rm nn}[l=l(k)].
\end{equation}
In Fig.~\ref{fig:analysis_topology}(h), we plot $k_{\rm nn}(k)$ against $k[l(k)]$ to find our approximated solution to be comparable with the simulation results.

\subsection{Intensity-related quantities}

\begin{figure}[!t]
    \includegraphics[width=\columnwidth]{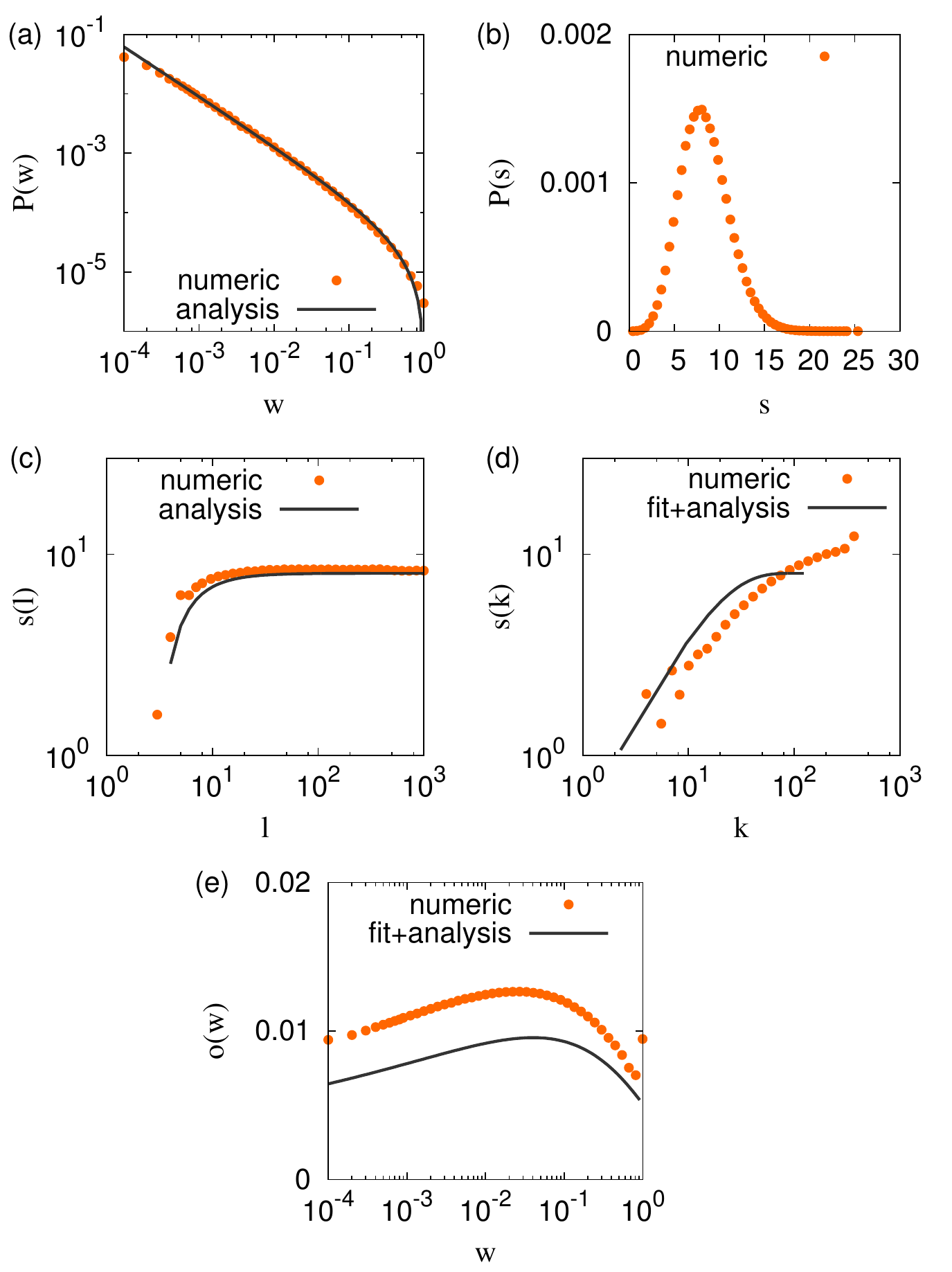}
    \caption{(Color online) Simulation results of intensity-related quantities for the same parameter values as in Fig.~\ref{fig:analysis_topology} with $\gamma=1.5$ and $w_0=1$ (circles) are compared to the analytical results (solid lines).}
    \label{fig:analysis_weight}
\end{figure}

Here we calculate local network quantities related to the intensity. The link weight distribution can be obtained as follows:
\begin{equation}
    \label{eq:define_Pw}
    P(w) \equiv \tfrac{1}{L}\sum_{g=g_0}^{g_{\rm max}} P_g(w) L_g
    \approx \tfrac{1}{L}\int_{g_0}^{g(w)} \tfrac{n_g g(g-1)p(g)}{2w(g)}dg,
\end{equation}
where $g(w)$ is the inverse of $w(g)$ in Eq.~(\ref{eq:model_wg}), i.e.,
\begin{equation}
    g(w)=g_0\left(\tfrac{w}{w_0}\right)^{-\frac{1}{\gamma}}.
\end{equation}
Then one obtains 
\begin{equation}
    \label{eq:Pw_approx}
    P(w)\approx \tfrac{g_0^{\beta-\gamma}CA}{2Lw_0}\left\{ H^{3,1}_{1,-1}[g_0,g(w)] -H^{2,1}_{1,-1}[g_0,g(w)]\right\}.
\end{equation}
This analytical result compares favorably with the simulation results in Fig.~\ref{fig:analysis_weight}(a). For relatively large $w$, we can get a simple scaling form from Eq.~(\ref{eq:Pw_approx}) as follows:
\begin{equation}
    P(w)\sim w^{-\frac{3-\alpha-\beta+\gamma}{\gamma}}.
    \label{eq:Pw_scaling}
\end{equation}
Thus, in order for $P(w)$ to decrease for large $w$, the following condition must be satisfied:
\begin{equation}
    \gamma>\alpha+\beta-3.
    \label{eq:decreasingPw}
\end{equation}
This condition narrows down the parameter region for stylized facts, as depicted in Fig.~\ref{fig:activity_alphagamma}(e).

Next, we obtain the expected strength in two forms: $s(k)$ and $s(l)$. We obtain $s(l)$ as
\begin{equation}
    \label{eq:define_sk}
    s(l) \equiv \sum_{g=g_0}^{l} \tfrac{N_g}{N} \langle k\rangle_g \langle w(g)\rangle,
\end{equation}
where $\langle w(g)\rangle \equiv \int_0^\infty wP_g(w)dw =\frac{w(g)}{2}$. One then obtains
\begin{equation}
    \label{eq:sk_approx}
    s(l)\approx \tfrac{g_0^{\beta+\gamma}w_0CA}{2N}\left[ H^{3,1}_{1,1}(g_0,l)- H^{2,1}_{1,1}(g_0,l)\right].
\end{equation}
This analytical result is compared with the simulation results in Fig.~\ref{fig:analysis_weight}(c). In most cases of our simulations, we use parameter values satisfying that $\alpha+\beta+\gamma>3$. It implies that $s(l)$ is approaching a constant from below, rather than increasing indefinitely. For the calculation of $s(k)$, we assume that 
\begin{equation}
    s(k)\equiv s[l=l(k)].
\end{equation}
In Fig.~\ref{fig:analysis_weight}(d), we plot $s(k)$ against $k[l(k)]$ to find our approximated solution to be comparable with the simulation results, but with systematic discrepancy.

Then, one can get the strength distribution $P(s)$ from the identity $P(s)ds=P(k)dk$ using the above $s(k)$. The peaked $P(k)$ combined with the increasing $s(k)$ leads to the peaked $P(s)$, as evidenced by the simulation results in Fig.~\ref{fig:analysis_weight}(b). The peak turns out to be around at the average strength $\langle s\rangle\approx \langle k\rangle \langle w\rangle\approx 8.4$.

Finally, we discuss the behavior of the neighborhood overlap $o(w)$. Once a link with weight $w$ is given, the link must have been created in a layer $g$ with $g_0\leq g\leq g(w)$. The probability $q_{w,g}$ that the link of weight $w$ is created in the layer $g$ reads
\begin{equation}
    q_{w,g} \equiv \tfrac{L_g/w(g)}{\sum_{g'=g_0}^{g(w)} L_{g'}/w(g')},
\end{equation}
where $\tfrac{1}{w(g)}$ comes from $P_g(w)$. If the link $ij$ is created in the layer $g$, nodes $i$ and $j$ have $(g-2)p(g)^2$ common neighbors on average, leading to the expected number of common neighbors as
\begin{eqnarray}
    &&\langle e\rangle_w \equiv \sum_{g=g_0}^{g(w)} q_{w,g}(g-2)p(g)^2.  \label{eq:e_w}\\
    &&\approx \tfrac{ g_0^{2\beta}\{ H^{4,1}_{3,-1}[g_0,g(w)] -3H^{3,1}_{3,-1}[g_0,g(w)] +2H^{2,1}_{3,-1}[g_0,g(w)]\}}{ H^{3,1}_{1,-1}[g_0,g(w)] -H^{2,1}_{1,-1}[g_0,g(w)]}.
\end{eqnarray}
Next, we take an approximation of both $k_i$ and $k_j$ in Eq.~(\ref{eq:o_ij}) as $k_{\geq}[g(w)]$ in Eq.~(\ref{eq:k_z_fit}) to finally obtain 
\begin{equation}
    o(w)\approx \tfrac{ \langle e\rangle_w }{ 2k_{\geq}[g(w)] -2 -\langle e\rangle_w}. 
\end{equation}
For our choice of parameter values, this analytical result of $o(w)$ turns out to be increasing and then decreasing with $w$, but with systematic discrepancy with the simulation results as shown in Fig.~\ref{fig:analysis_weight}(e). The increasing behavior of $o(w)$ for small $w$ can be understood by considering a simple scaling form for large $g(w)$. Since $\alpha+\beta>3$ and $3-\alpha-3\beta+\gamma>0$ for our choice of exponent values, we get
\begin{equation}
    o(w)\sim \tfrac{ \langle e\rangle_w }{k_{\geq}[g(w)]} \sim w^{\frac{2\beta-1}{\gamma}}.
\end{equation}
Thus, the increasing behavior of $o(w)$ for small $w$ is realized for $\beta>\frac{1}{2}$, and enhanced by the smaller $\gamma$. This is because the relatively large $\beta$ reduces the possibility of having common neighbors in large communities, which leads to smaller neighborhood overlaps for weaker links. However, this argument does not apply to the very strong links as the value of $p(g)$ for small $g$ is still high for large $\beta$. This may account for the decreasing behavior of $o(w)$ for very large $w$ that has been empirically observed in some datasets~\cite{Onnela2007Structure, Noka2016Comparative}.

\section{Conclusions}\label{sect:conclusion}

Despite a number of empirical findings for social networks, the multi-channel weighted social (MWS) network 
has never been comprehensively identified, prompting us to ask a series of important questions: What does the MWS network look like? Then how can it be parsimoniously modeled or reproduced? As for the first question, we expect that the structure of the MWS network is reflected in the empirical findings from partial datasets of the network. For this, we have found several commonly observed features or stylized facts in various datasets of social networks, such as broad distributions of local network quantities, existence of communities, assortative mixing, and intensity-topology correlations. These stylized facts are listed in Table~\ref{table:summary}. Among them, the overall decreasing degree and strength distributions are not fully consistent with common sense. In particular, we expect the degree and strength distributions to be peaked~\cite{Murase2015Modeling, Kertesz2016Multiplex}.

As for the parsimonious modeling of the MWS network, we have devised the community-based static model for reproducing the stylized facts with the expected peaked behavior of degree and strength distributions. For our model, we have randomly assigned a number of communities to a given set of isolated nodes using three assumptions, such that (i) the size $g$ of each community is drawn from a power-law distribution with the exponent $\alpha$, (ii) the link density for the community of size $g$ is given as a decreasing function of $g$, controlled by the power-law exponent $\beta$, and (iii) the characteristic weight of links created in the community of size $g$ is given as a decreasing function of $g$, controlled by the power-law exponent $\gamma$. With these few assumptions about communities, realistic social networks have successfully been reproduced such that they show almost all stylized facts for a wide range of parameter space of $(\alpha,\beta,\gamma)$. Note that our assumptions are reasonable, yet they could be deduced from more fundamental mechanisms for the creation, aging, and severance of links. 

We could obtain some analytical results for local network quantities that compare favorably with the numerical results. It is because our model can be interpreted as the aggregate of random uniform hypergraphs with various degrees. Such analytic results indeed provide deeper understanding of the consequences of the model. More importantly, thanks to the simplicity and explicitness of our model, we expect it to serve as a general reference system for further research in the direction of investigating the structure and dynamics of the MWS network, as well as the dynamical processes taking place in the MWS network. In addition, once we better understand how the sampling relates the MWS network to the partial observations, we can better translate the conclusions drawn from partial datasets into those for the MWS network.

We remark that we have ignored the known correlations such as those due to the geographical constraint~\cite{Onnela2011Geographic} and/or demographic information~\cite{Jo2014Spatial}. These correlations can be incorporated for more realistic modeling of the MWS network. For example, since the geographical distance between a pair of nodes is found to be negatively correlated with the link probability between them~\cite{Onnela2011Geographic}, geographically close nodes can be chosen when assigning communities to nodes, as done in Ref.~\cite{Murase2014Multilayer}.

Finally, we briefly discuss the rank curve analysis as one of the stylized facts for social networks. How individuals distribute their limited resources like time to their neighbors is also indicative in characterizing the social networks at the individual level. The rank curve of a node is defined by the weights of links involving the node in a descending order. The layered structure in rank curves has been claimed~\cite{Dunbar2011Constraints, MacCarron2016Calling}, while the exponential or power-law functional forms have been successfully used for fitting empirical rank curves~\cite{Song2013Connections, Saramaki2014Persistence}. In fact, there is a chance to obtain the layered or stepwise rank curves from our model by assuming that all links in the same layer have the same weight, e.g., by using $P_g(w)=\delta[w-w(g)]$ in Eq.~(\ref{eq:Pgw_delta}). However, this functional form for $P_g(w)$ requires too strict conditions for the parameter values, as discussed in Appendix~\ref{appendix:Pgw}. As the functional form of rank curves is yet inconclusive, we leave this issue for a future work.

\begin{acknowledgments}
H.-H.J. acknowledges financial support by Basic Science Research Program through the National Research Foundation of Korea (NRF) grant funded by the Ministry of Education (2015R1D1A1A01058958) and the framework of international cooperation program managed by the National Research Foundation of Korea (NRF-2016K2A9A2A08003695). 
Y.M. appreciates hospitality at Aalto University and acknowledges support from CREST, JST. 
J.T. thanks for financial support of Aalto AScI internship programme. 
J.K. acknowledges support from EU Grant No. FP7 317532 (MULTIPLEX).
K.K. acknowledges support from Academy of Finland's COSDYN project (No. 276439) and EU's Horizon 2020 FET Open RIA 662725 project IBSEN.
This project was partly supported by JSPS and NRF under the Japan-Korea Scientific Cooperation Program. Partial support by OTKA, K112713 is also acknowledged. The systematic simulations in this study were assisted by OACIS~\cite{Murase2014Tool}.  
\end{acknowledgments}

\appendix

\section{Alternative forms of $P_g(w)$}\label{appendix:Pgw}

In the main text, we studied the case with the uniform distribution for $P_g(w)$. Here we discuss two alternative forms of $P_g(w)$. The first case is the delta function as
\begin{equation}
    \label{eq:Pgw_delta}
    P_g(w)=\delta[w-w(g)],
\end{equation}
with the same $w(g)=w_0(\tfrac{g_0}{g})^{\gamma}$ as in Eq.~(\ref{eq:model_wg}). We then find that the parameter range for stylized facts is strongly limited. For example, we obtain the weight distribution as $P(w)\sim w^{-\frac{2-\alpha-\beta}{\gamma}}$. It implies that for decreasing $P(w)$ one must have $\alpha+\beta<2$, which is hardly realistic, considering the empirical values of these exponents~\cite{Palla2005Uncovering, Ahn2010Link, Liu2012Social, Grabowicz2012Social, Hric2014Community}. As for the second case, we can study the exponential distribution as follows:
\begin{equation}
    P_g(w)=\tfrac{1}{w(g)}e^{-w/w(g)}.
\end{equation}
We find the same scaling behavior for large $w$ as in Eq.~(\ref{eq:Pw_scaling}). Thus, the advantage of this form is only marginal compared with our choice of the uniform distribution.

\section{Overlapping communities}\label{appendix:overlap}

In order to consider the case of overlapping communities, we introduce an overlapping probability $q_{gg'm}$ that two communities of size $g$ and $g'$ share $m$ nodes with $1\leq m\leq \min\{g,g'\}$. Here we assume that $p(g)=1$ for all $g$, which provides an upper bound of $q_{gg'm}$. One obtains
\begin{eqnarray}
    \nonumber
  q_{gg'm}&\equiv&\frac{{N \choose m}{N-m \choose g-m}{N-g\choose g'-m}}{{N \choose g}{N\choose g'}}\\
    \nonumber
  &=&\tfrac{g!g'!}{m!(g-m)!(g'-m)!} \tfrac{(N-g)!(N-g')!}{N!(N-g-g'+m)!}\\
  &\approx& \tfrac{m!}{N^m}{g\choose m}{g'\choose m}.
\end{eqnarray}
We get the basic quantities as
\begin{eqnarray}
 q_{gg1}\approx\tfrac{g^2}{N},\ && q_{gg2}\approx\tfrac{[g(g-1)]^2}{2N^2},\\
 q_{gg'1}\approx\tfrac{gg'}{N},\ && q_{gg'2}\approx\tfrac{g(g-1)g'(g'-1)}{2N^2}.
\end{eqnarray}
Provided that $N\gg g$, $q_{gg'2}$ is found to be mostly negligible. It implies that it is very unlikely to find a pair of nodes belonging to multiple communities, whether communities have the same size or not, i.e., whether they are in the same layer or not. The average number of links in the layer $g$ can be obtained as
\begin{eqnarray}
    \nonumber\label{eq:define_Lg_exact}
    L_g&=& n_g\tfrac{g(g-1)}{2} p(g)- \tfrac{n_g(n_g-1)}{2} \sum_{m=2}^g \tfrac{m(m-1)}{2}p(g) q_{ggm}\\
    &\approx& n_g\tfrac{g(g-1)}{2} p(g) - \tfrac{n_g(n_g-1)}{2} p(g) \tfrac{[g(g-1)]^2}{2N^2}.
\end{eqnarray}
For the second line, we have taken only the dominant term in the summation.

\begin{figure}[!t]
    \includegraphics[width=\columnwidth]{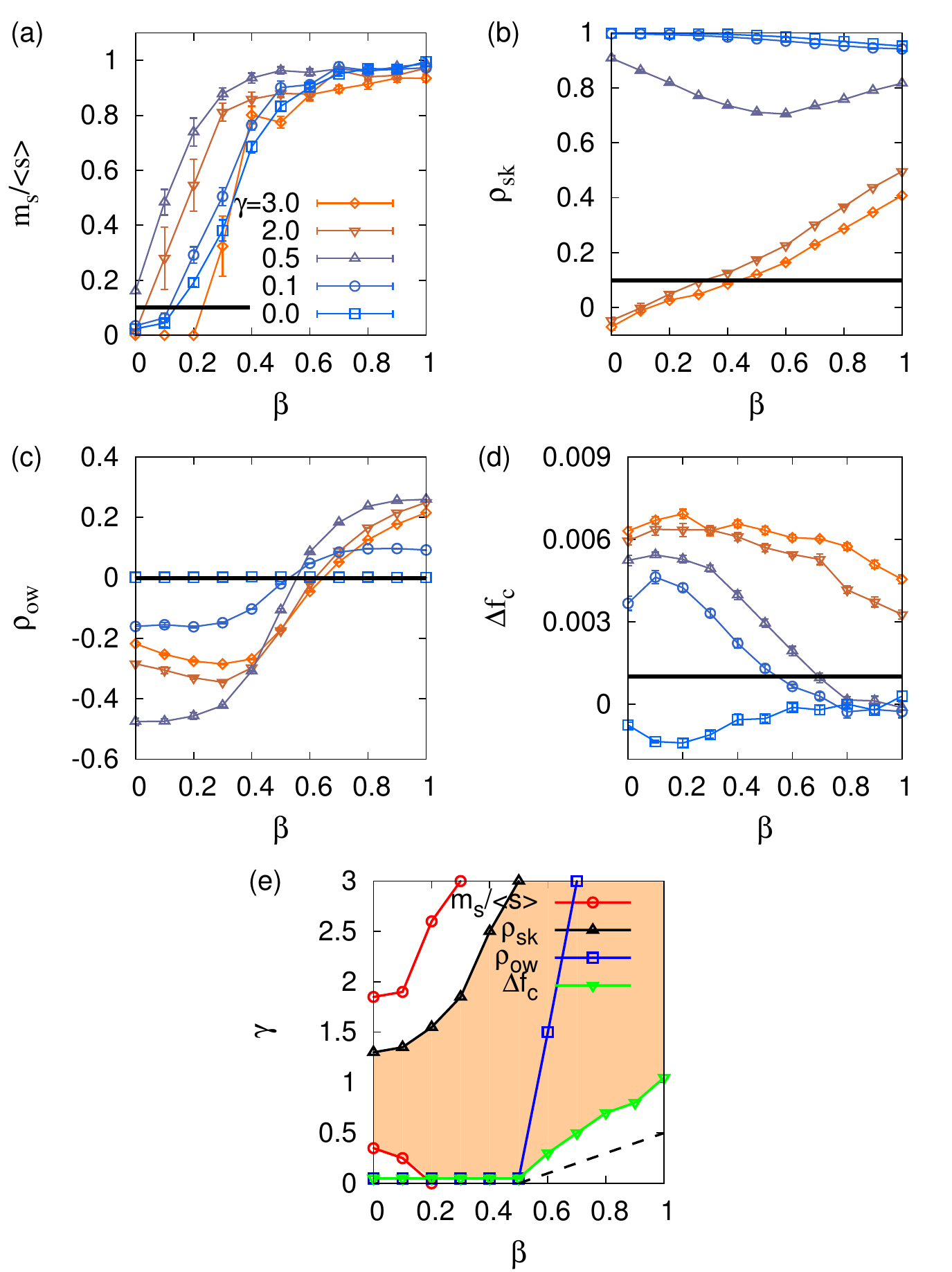}
    \caption{(Color online) Effects of $\beta$ and $\gamma$ on intensity-related properties for a given $\alpha=2.5$: (a) $\frac{m_s}{\langle s\rangle}$, (b) $\rho_{sk}$, (c) $\rho_{ow}$, and (d) $\Delta f_c$, with corresponding threshold values (black lines) for stylized facts. (e) Assuming that the intensity-related stylized facts are reproduced when $\frac{m_s}{\langle s\rangle}>0.1$, $\rho_{sk}>0.1$, $\rho_{ow}>0$, and $\Delta f_c>0.001$, we barely find the parameter region surrounded by four curves from (a--d). If the condition for positive $\rho_{ow}$ is relaxed, then we have the wide range of parameter space (shaded area) for the stylized facts, i.e., for large $\beta$ and intermediate $\gamma$. Here the results have been averaged over $10$ networks generated using $N=3\cdot 10^4$, $\langle k\rangle=100$, $g_0=3$, $g_{\rm max}=10^3$, and $w_0=1$. The dashed line, $\gamma=\alpha+\beta-3$, indicates the criterion for the decreasing behavior of $P(w)$ obtained from Eq.~(\ref{eq:decreasingPw}).}
    \label{fig:activity_betagamma}
\end{figure}

We can also calculate the exact number of nodes chosen for communities of size $g$, denoted by $N_g$, by considering the effect by overlapping communities. The maximum of $N_g$ is $gn_g$ as $n_g$ communities are assigned with $g$ nodes for each community. $N_g$ can be smaller than $gn_g$ due to two factors: (i) Some nodes may be isolated in the layer $g$ if $p(g)<1$. (ii) Some nodes may belong to multiple communities. For the former, the probability of a node being isolated is $[1-p(g)]^{g-1}$, provided that this node could belong to only one community. For the latter, the dominant case is when two communities overlap over one node, whose probability is $q_{gg1}$. In sum, one obtains
\begin{equation}
    N_g\approx gn_g\{1-[1-p(g)]^{g-1}\}-\tfrac{n_g(n_g-1)}{2}\tfrac{g^2}{N}.
\end{equation}
We then get the average degree only for nodes chosen for communities in the layer $g$ as
\begin{equation}
    \langle k\rangle_g = \tfrac{2L_g}{N_g}\approx \tfrac{(g-1)p(g)-
        \frac{1}{2N^2}(n_g-1)p(g)g(g-1)^2}{1-[1-p(g)]^{g-1}-\frac{1}{2N}(n_g-1)g}.
\end{equation}
If $p(g)\ll 1$ and $N$ is sufficiently large, one obtains $\langle k\rangle_g\approx 1$, implying that links are mostly isolated. 

\section{Effects of $\beta$ and $\gamma$ for a fixed $\alpha=2.5$}\label{appendix:betagamma}

We numerically obtain the intensity-related quantities, i.e., $\frac{m_s}{\langle s\rangle}$, $\rho_{sk}$, $\rho_{ow}$, and $\Delta f_c$, by varying $\beta$ and $\gamma$ while keeping $\alpha=2.5$, for which it is expected to show the topological stylized facts for $0.1<\beta<0.6$. As shown in Fig.~\ref{fig:activity_betagamma}, we apparently find a wide region in the parameter space of $(\beta, \gamma)$, where all intensity-related stylized facts are reproduced, i.e., relatively large $\frac{m_s}{\langle s\rangle}$, $\rho_{sk}$, and $\Delta f_c$, as well as positive $\rho_{ow}$, by using the same threshold values for these quantities as discussed in the main text. However, as the topological stylized facts are reproduced for $0.1<\beta<0.6$, the parameter region for the intensity-related stylized facts must be narrow. Thus, if the condition for positive $\rho_{ow}$ is relaxed again, we have the wide range of the parameter space for the stylized facts to be reproduced, i.e., for large $\beta$ and intermediate $\gamma$. This region is depicted as shaded in Fig.~\ref{fig:activity_betagamma}(e). 

\section{Standard deviation for the degrees}\label{appendix:std_degree}

In order to take into account the correlation between links imposed by the communities in the degree distribution, we separate nodes chosen for communities from those not chosen in each layer $g$. Then $P_g(k)$ consists of two parts, one for nodes chosen for communities and the other for those not chosen:
\begin{equation}
    \label{eq:Pgk}
    P_g(k)=\left(1-\tfrac{N_g}{N}\right)\delta_{k,0}+ \tfrac{N_g}{N} {g-1\choose k} p(g)^k[1-p(g)]^{g-1-k},
\end{equation}
where $k=0,\cdots,g-1$. Note that isolated nodes are either those not chosen for communities or those chosen but left with no links when $p(g)<1$. The calculation of $P(k)$ from the above $P_g(k)$ is not trivial. In order to calculate the standard deviation for the degrees, we assume that
\begin{equation}
    P_g(k)=\left(1-\tfrac{N_g}{N}\right)\delta(k)+ \tfrac{N_g}{N}\delta(k-\langle k\rangle_g).
\end{equation}
One obtains the variance for the layer $g$ as
\begin{equation}
    \sigma^2_g = \tfrac{N_g}{N}\left(1-\tfrac{N_g}{N}\right) \langle k\rangle^2_g,
\end{equation}
leading to the variance for the entire network as
\begin{eqnarray}
    && \sigma^2 \equiv \sum_{g=g_0}^{g_{\rm max}} \sigma^2_g \approx \int_{g_0}^{g_{\rm max}} \sigma^2_g dg\\
    &&=\tfrac{g_0^{2\beta}CA}{N}\left[ H^{4,1}_{2,0}(g_0,g_{\rm max}) -2H^{3,1}_{2,0}(g_0,g_{\rm max})+ H^{2,1}_{2,0}(g_0,g_{\rm max}) \right]\nonumber\\
    &&- \tfrac{g_0^{2\beta}(CA)^2}{N^2}\left[ H^{5,2}_{2,0}(g_0,g_{\rm max}) -2H^{4,2}_{2,0}(g_0,g_{\rm max})+ H^{3,2}_{2,0}(g_0,g_{\rm max}) \right],\nonumber\\
    \label{eq:std_degree}
\end{eqnarray}
where $H^{u,v}_{m,n}$ is defined in Eq.~(\ref{eq:H}).

%

\end{document}